\documentclass[preprint,prb,amsmath,amssymb,amsfonts,superscriptaddress,floatfix,showpacs]{revtex4}
\usepackage{graphicx}% Include figure files
\usepackage{dcolumn}% Align table columns on decimal point
\usepackage{bm}% bold math
\usepackage{rotating} % rotate figure

\begin{document}

%\title{All-electron GW approximation self-energy based on the PAW method}
%\title{All-electron GW approximation based on the PAW method}
\title{
Electronic structure of two-dimensional crystals from ab-initio theory
}
\date{\today}
\author{S.~Leb\`egue}
\affiliation{
Laboratoire de Cristallographie et de Mod\'elisation des Mat\'eriaux Min\'eraux et Biologiques, 
UMR 7036, CNRS-Universit\'e Henri Poincar\'e, B.P. 239, F-54506 Vandoeuvre-l\`es-Nancy, France
}
\author{O.~Eriksson}
\affiliation{
Department of Physics and Materials Science, University of Uppsala, SE-75121 Uppsala, Sweden 
}

\begin{abstract}
	We report on ab-initio calculations of the two-dimensional systems MoS$_2$ and NbSe$_2$,
  which recently were synthesized. We find that two-dimensional MoS$_2$ is a semiconductor with a gap which is rather close 
  to that of the three dimensional analogue, and that NbSe$_2$ is a metal, which is similar to the three dimensional analogue of this compound.
	  We further computed the
	   electronic structure of the two-dimensional hexagonal (graphene like) lattices of Si and Ge, and compare
	   them with the electronic structure of graphene. 
	    It is found that the properties related to the Dirac cone do not appear in the case of 
	     two-dimensional hexagonal germanium, which is metallic, contrary to two-dimensional hexagonal silicon,
	      which has an electronic structure very similar to the one of graphene, making them
	       possibly equivalent.
\end{abstract}

\pacs{71.15.Mb, 71.20.-b}% PACS, the Physics and Astronomy
%                             % Classification Scheme.
%\keywords{Suggested keywords}%Use showkeys class option if keyword
                              %display desired
%\preprint{APS/123-QED}

\maketitle
%%%%%%%%%%%%%%%%%%%%%%%%%%%%%%%%%%%%%%%%%%%%%%%%%%%%%%%%%%
%%%%%%%%%%%%%%%%%%%%%%%%%%%%%%%%%%%%%%%%%%%%%%%%%%%%%%%%%%
%%%%%%%%%%%%%%%%%%%%%%%%%%%%%%%%%%%%%%%%%%%%%%%%%%%%%%%%%%

\section{\label{sec:one} Introduction}

During several years, it was believed that the existence of free-standing two dimensional crystals was impossible, because they
would be unstable and ultimately return to a three dimensional object\cite{perl,landau}. However, it appeared recently that isolated sheets
of graphene could be obtained by mechanical exfoliation of a graphite crystal, which therefore proves the predictions of Ref.\onlinecite{perl,landau} to be inaccurate\cite{electric}.
  The experimental accomplishment of synthesizing two-dimensional crystals has lead to the emergence of a truly new physics, since the particular properties of graphene can be considered 
  as a bridge between quantum electrodynamics (QED) and condensed matter physics\cite{misha1}. Indeed, near the $K$ point of the Brillouin zone (BZ), 
   the one-particle energy dispersion is linear with the momentum, and therefore the corresponding quasiparticles
   could be described by a Dirac-like Hamiltonian.  Then, QED properties can be studied by investigating the electronic structure of graphene.

Later on, it was shown\cite{PNAS2005} that the same technique could be used to obtain other compounds, opening the path to the 
 investigation of a large number of two dimensional crystals. However, it seems that most of the experimental and
  theoretical efforts in this area are still focusing exclusively on graphene, primarily because of the high quality
   of samples that could be obtained. In the present paper, we investigate by means of ab-initio calculations the
    electronic structure of MoS$_2$ and NbSe$_2$, as they have been synthesized experimentally\cite{PNAS2005}, as well
     as hypothetical two dimensional hexagonal crystals (with a graphene like structure) made of Si and Ge.

\section{\label{sec:two} Computational details}

To perform the calculations, we have used density functional
theory (DFT) \cite{Hohenberg,Sham} as implemented in the code VASP (Vienna Ab-initio simulation package)\cite{vasp2,vasp},
 within the framework of the PAW (projector augmented waves) method \cite{Bloechl}.
 The Local Density Approximation\cite{ceperley} (LDA) as well as the Perdew Burke Ernzerhof\cite{PBE} variant of the generalized gradient approximation (GGA)
 were used for the exchange-correlation potential.
A cut-off of $500$ eV was used for the plane-wave expansion of the wave function to converge the relevant quantities.
 For Brillouin zone integrations, a mesh of $40 \times 40 \times 1 $  $k$-points\cite{Monkhorst} was used for MoS$_2$ and NbSe$_2$,
 while a mesh of $20 \times 20 \times 3 $ was sufficient to describe two-dimensional silicon and germanium.

 Concerning the crystal structure MoS$_2$ and NbSe$_2$ we used the experimental bulk values
  since it was noticed in Ref. \onlinecite{PNAS2005} that $2$D crystals in these cases remains very close
   to their 3D parents. In this case, GGA was used. For the two dimensional hexagonal crystals made of Si and Ge,
    we optimized the lattice parameter using either the LDA or the GGA.
    Moreover, it appeared that the band structure obtained using either LDA or GGA were very similar, so
     we have chosen to show only results corresponding to LDA.
    For all studies here, the 'c' parameter was taken to be large enough to ensure that no interaction remains
     between layers, making them effectively isolated $2$D objects.

\section{\label{sec:three} The electronic structure of two-dimensional MoS$_2$ and NbSe$_2$}

\subsection{\label{sec:three-mo} MoS$_2$}

There is a big interest in transition metal dichalcogenides,
 since due to their layered structure, they have extremely anisotropic properties,
  and therefore an intercalation process is easy to conduct. In particular, MoS$_2$
   is used in the technology of Li batteries.
   To investigate purely two-dimensional MoS$_2$, we have used the bulk lattice parameter (a $= 3.16$ \AA), 
   as reported in Ref \onlinecite{Hussain}. Since reduced dimensionality can sometimes leads to magnetic
    behavior in systems which are not magnetic in bulk, we checked the possibility to have a spin-polarized
     ground-state, but it was found that two-dimensional MoS$_2$ remains in a non-magnetic state.

\begin{figure}[h]
\includegraphics*[angle=0,width=0.7\textwidth]{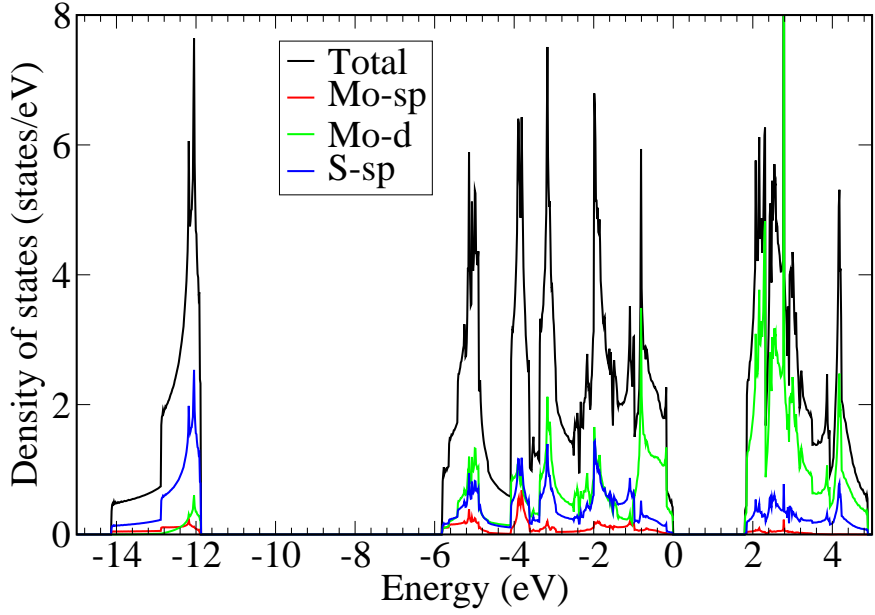}
\caption{
The density of states of two-dimensional MoS$_2$.
The Fermi level is put at zero eV.
\label{fig:dos-mos2} 
}
\end{figure}

Our calculated DOS is presented in Fig \ref{fig:dos-mos2}. As for three-dimensional MoS$_2$, the two-dimensional variant of this compound
is semiconducting: the bands on each side of the band gap are derived mainly from the Mo-d states, which is what a calculation of three dimensional MoS$_2$ also shows\cite{Mattheiss}. Also, a large gap
 exists in the occupied states, separating the S-s states (between -14 and -12 eV) and the hybridized Mo and S-p states (between -6 and 0 eV).
\begin{figure}[h]
\includegraphics*[angle=0,width=0.7\textwidth]{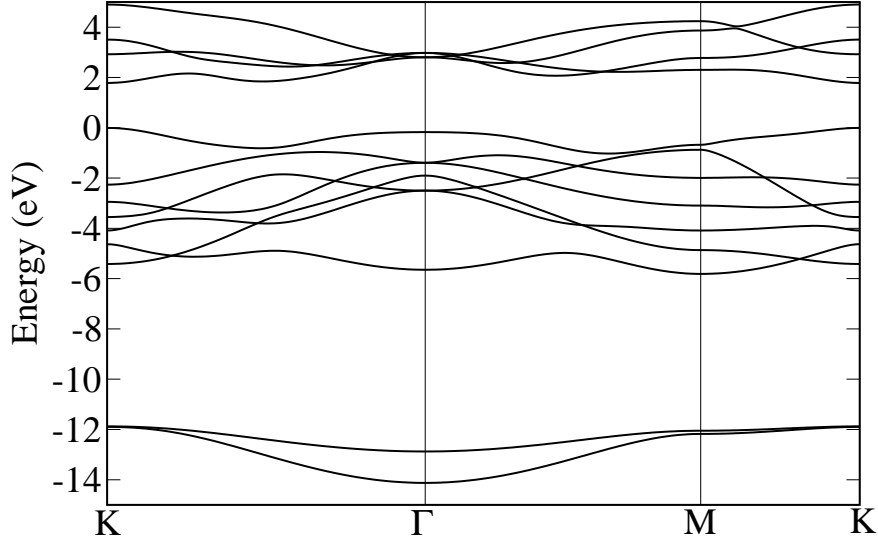}
\caption{
\label{fig:bnds-mos2} 
The band structure of two-dimensional MoS$_2$.
The Fermi level is put at zero eV.
}
\end{figure}
In Fig. \ref{fig:bnds-mos2}, we present the energy bands of two-dimensional MoS$_2$.
We see again a lot of similarities with three dimensional MoS$_2$: the bands around the band-gap are relatively
 flat, as expected from the 'd' character of the electron states at these energies.
The band gap, which has a value of 1.78 eV, is direct, and occurs at the high-symmetry point $K$, whereas for three dimensional MoS$_2$ 
 the band gap is indirect\cite{Mattheiss}.

\subsection{\label{sec:three-nb} NbSe$_2$}

As for MoS2, the NbSe$_2$ structure is strongly layered, with each Nb layer being sandwiched between two Se layers, where weak van der Waals forces are holding the whole geometry together. NbSe$_2$ is known to be a prototype to study charge density waves (CDW),
and is also a superconductor. Here we focus only on the electronic structure of two-dimensional NbSe$_2$ and a lattice constant of $3.45$ \AA\cite{corcoran} was used to perform the calculation.

\begin{figure}[h]
\includegraphics*[angle=0,width=0.7\textwidth]{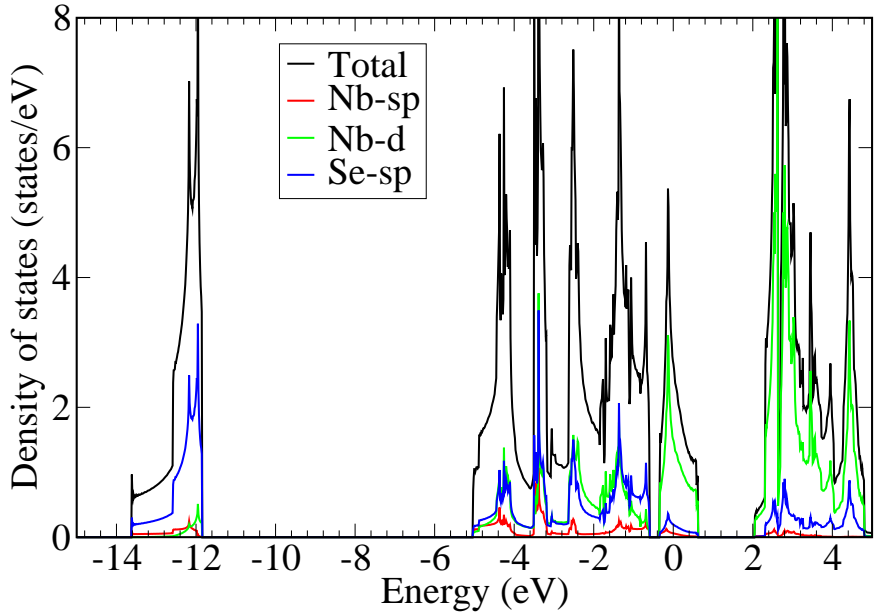}
\caption{
The density of states of two-dimensional NbSe$_2$. 
The Fermi level is put at zero eV.
\label{fig:dos-nbse2} 
}
\end{figure}
\begin{figure}[h]
\includegraphics*[angle=0,width=0.7\textwidth]{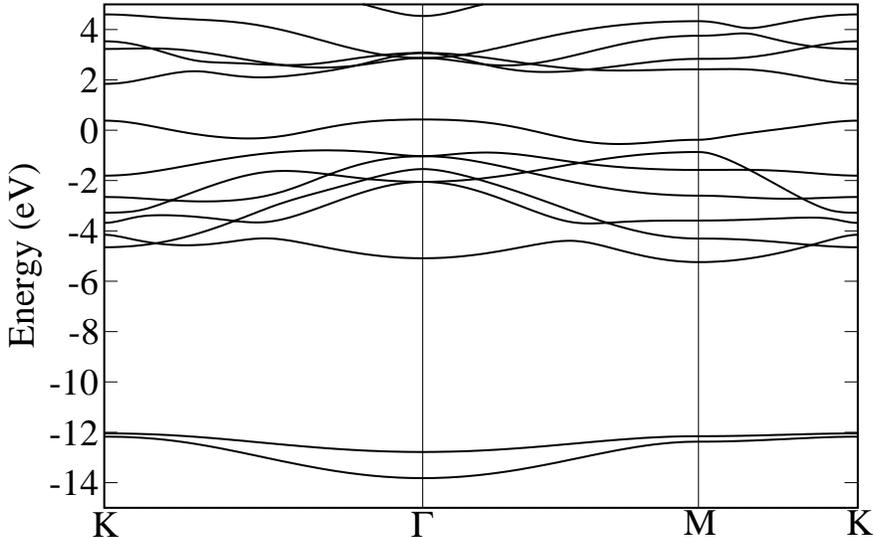}
\caption{
The band structure of two-dimensional NbSe$_2$.
The Fermi level is put at zero eV.
\label{fig:bnds-nbse2} 
}
\end{figure}

As for MoS$_2$, we have checked the existence of a magnetic solution, since the non spin-polarized DOS (Fig. \ref{fig:dos-nbse2})
 is very peaked at the Fermi level, but we found again a non-magnetic ground state. However, contrary to  MoS$_2$, two dimensional NbSe$_2$
  is metallic, just like the three dimensional version of this compound. The DOS of two-dimensional NbSe$_2$ shows a lot of similarities
  with the one of MoS$_2$, with a large hybridization between Nb and Se states. In addition, the band state which is pinned at the Fermi level
   is derived primarily from Nd d-orbitals, and this band state becomes separated in energy from all other states (Fig. \ref{fig:bnds-nbse2}).

\section{\label{sec:four} The electronic structure of two-dimensional Si and Ge}

With four 'sp' electrons in the valence band, silicon and germanium (possibly together with BN) are probably the closest
to carbon from a chemical point of view. Therefore, it is of interest to see if in their
two-dimensional form they could present some similarities with graphene, and we have investigated this possibility here.

\begin{table}
\begin{center}
\begin{tabular}{|l|r|r|}
\hline
\multicolumn{1}{|c|}{System } &
\multicolumn{1}{|c|}{ LDA } &
\multicolumn{1}{|c|}{ GGA}
\\\hline\hline
Si  &  3.860 \AA  &  3.901 \AA \\\hline
Ge  &  4.034 \AA  &  4.126 \AA \\\hline
\end{tabular}
\end{center}
\caption{
Lattice parameters of hexagonal Si and Ge computed with either LDA or GGA.
\label{tab:lattice} 
}
\end{table}

\begin{figure}[h]
\includegraphics*[angle=0,width=0.7\textwidth]{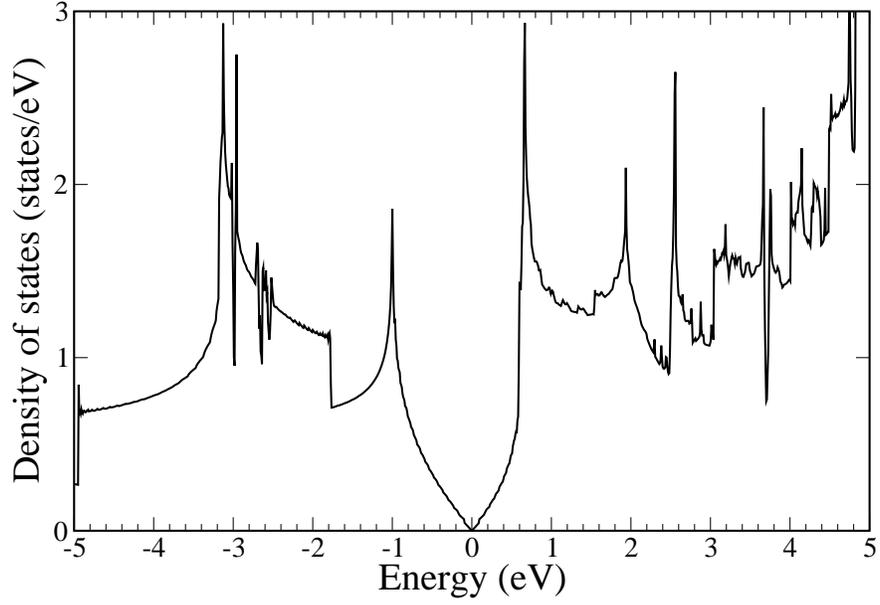}
\caption{
The density of states of two-dimensional Si.
The Fermi level is put at zero eV.
\label{fig:dos-si} 
}
\end{figure}

\begin{figure}[h]
\includegraphics*[angle=0,width=0.7\textwidth]{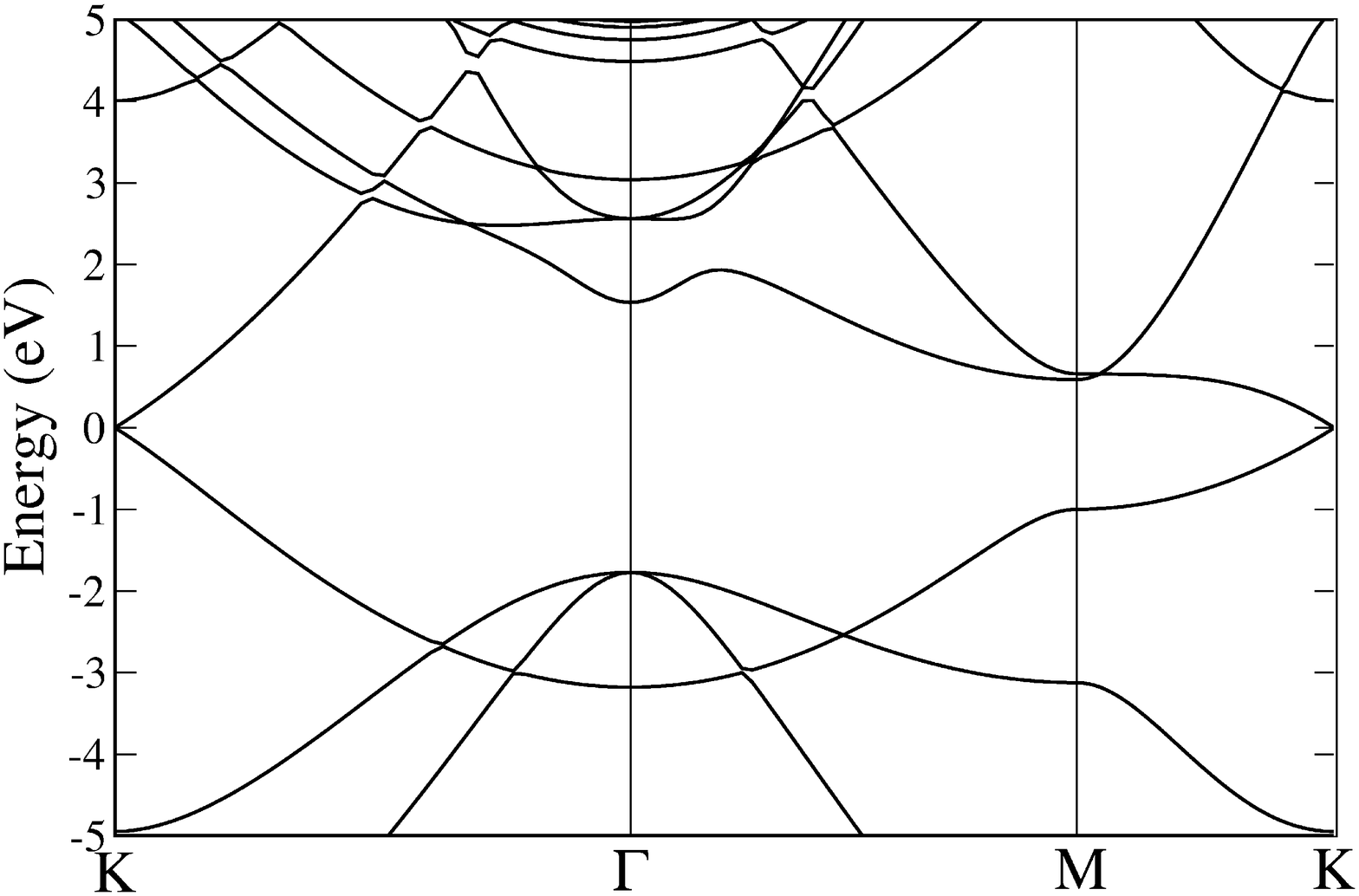}
\caption{
The band structure of two-dimensional Si.
The Fermi level is put at zero eV.
\label{fig:bnds-si} 
}
\end{figure}

In Table \ref{tab:lattice}, we present our computed lattice parameters (see computational section) for hexagonal Si and Ge.
As expected, they are larger than the one of graphene ($2.46$ \AA) because of the larger radius
of Si and Ge in comparison with C. Also, because of the well-known overbinding of LDA, 
GGA gives lattice parameters which are larger than the ones given by LDA.

\begin{figure}[h]
\includegraphics*[angle=0,width=0.7\textwidth]{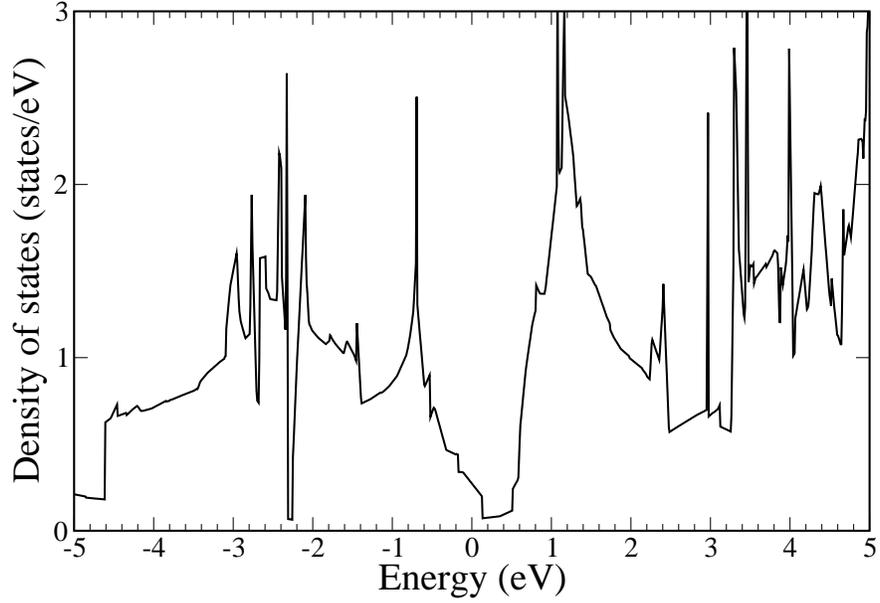}
\caption{
The density of states of two-dimensional Ge.
The Fermi level is put at zero eV.
\label{fig:dos-ge} 
}
\end{figure}

\begin{figure}[h]
\includegraphics*[angle=0,width=0.7\textwidth]{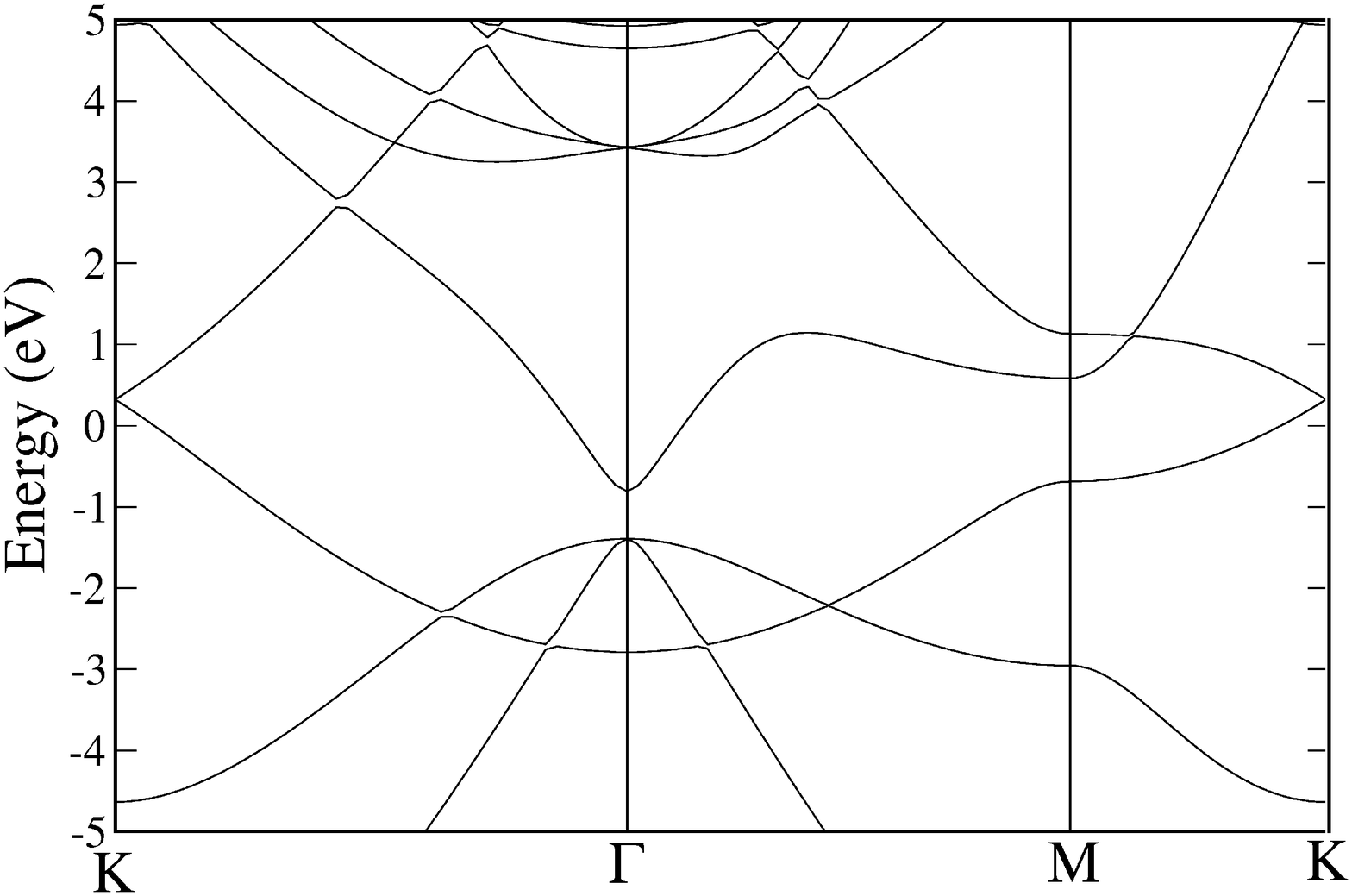}
\caption{
The band structure of two-dimensional Ge.
The Fermi level is put at zero eV.
\label{fig:bnds-ge} 
}
\end{figure}

We used these lattice parameters for the calculation of the corresponding densities of states and band structures.
As seen in Figs \ref{fig:dos-si} and \ref{fig:bnds-si}, two-dimensional silicon shows a lot of similarities
 with graphene: in particular the gap is also closing at the K point of the Brillouin zone, and the dispersion
  around this point is linear.
In contrast, two-dimensional germanium is quantitatively different from graphene:
 a conduction band is partially filled by electrons around the $\Gamma$ point and therefore
  the bands at the $K$ point are shifted up in energy, so that two-dimensional germanium is not
   a zero-gap material, but rather a poor metal. For three dimensional germanium,
    it is known that LDA/GGA gives a non band-gap behavior, and a more advanced technique
    like the GW approximation\cite{Hedin1,Hedin2,prbgw} is required to obtain a band gap. 
    However, our calculations with DFT show that two-dimensional germanium is much more metallic 
     than three-dimensional germanium, and therefore the metallic nature is here much more robust and
      is likely to be preserved even if a quasiparticle theory (like the GW approximation) would
       be used.

\section{\label{sec:five} Conclusion}

We have studied by means of ab-initio calculations the electronic structure of 
 two-dimensional MoS$_2$ and NbSe$_2$, as well as hypothetical graphene like structures of Si and Ge.
  We have found that two-dimensional silicon might from an electronic structure point of view be equivalent to graphene,
  with a linear dispersion of the electronic structure around the K-point. The possible advantage with Si in this regard is that it
probably is more easily interfaced with existing electronic devices and technologies. The obvious disadvantage is that sp$^2$ bonded
 Si is much less common than for C, and the synthesis of Si in a graphene like structure is extremely demanding and is likely to represent 
 a meta-stable material. The electronic structure of Ge in the graphene structure results in a metallic behavior, and the electronic structure 
  of two-dimensional MoS$_2$ and NbSe$_2$ is quite similar to that of the three dimensional 
 counterparts, with MoS$_2$ being a gapped system with a direct gap and NbSe$_2$ a metal.

\begin{acknowledgments}
 S.L. acknowledges financial support from ANR PNANO Grant ANR-
 06-NANO-053-02 and ANR Grant ANR-BLAN07-1-186138, as well as CINES/CCRT
  for computer time.
  O. E. is grateful to SNAC and VR for support and to a KoF initiative at Uppsala University.
\end{acknowledgments}

\end{document}